\documentclass[aps,prl,superscriptaddress,twocolumn]{revtex4-2}
\usepackage{graphics}
\usepackage{epsfig}
\usepackage{color}
\usepackage{stackrel}
\usepackage{amsfonts}
\usepackage{wrapfig}
\usepackage{pstricks}
\usepackage{pst-node}
\usepackage{bm}
\usepackage{dcolumn}
\usepackage{multirow}
\usepackage{epstopdf}
\usepackage{subfigure}
\usepackage{amssymb}
\usepackage{amsmath}
\usepackage{commath}
\usepackage{graphicx,bm}
\usepackage{verbatim}
\usepackage{booktabs}
\usepackage[para]{threeparttable}
\usepackage{etoolbox}
\usepackage{lipsum}

\begin{document}
\title{Hexcitons and oxcitons in monolayer WSe$_2$}
\author{Dinh~Van~Tuan}
\affiliation{Department of Electrical and Computer Engineering, University of Rochester, Rochester, New York 14627, USA}
\author{Su-Fei~Shi}
\affiliation{Department of Chemical and Biological Engineering, Rensselaer Polytechnic Institute, Troy, NY, 12180, USA}
\affiliation{Department of Electrical, Computer and Systems Engineering, Rensselaer Polytechnic Institute, Troy, NY, 12180, USA}
\author{Xiaodong~Xu}
\affiliation{Department of Physics, University of Washington, Seattle, Washington 98195, USA}
\affiliation{Department of Materials Science and Engineering, University of Washington, Seattle, Washington 98195, USA}
\author{Scott~A.~Crooker}
\affiliation{National High Magnetic Field Laboratory, Los Alamos, New Mexico 87545, USA}
\author{Hanan~Dery}
\altaffiliation{hanan.dery@rochester.edu}
\affiliation{Department of Electrical and Computer Engineering, University of Rochester, Rochester, New York 14627, USA}
\affiliation{Department of Physics and Astronomy, University of Rochester, Rochester, New York 14627, USA}
\begin{abstract}
In the archetypal monolayer semiconductor WSe$_2$, the distinct ordering of spin-polarized valleys (low-energy pockets) in the conduction band allows for studies of not only simple neutral excitons and charged excitons (i.e., trions), but also more complex many-body states that are predicted at higher electron densities. We discuss magneto-optical measurements of electron-rich WSe$_2$ monolayers, and interpret the spectral lines that emerge at high electron doping as optical transitions of 6-body exciton states (``hexcitons'') and 8-body exciton states (``oxcitons''). These many-body states emerge when a photoexcited electron-hole pair interacts simultaneously with multiple Fermi seas, each having distinguishable spin and valley quantum numbers. In addition, we identify the energies of primary and satellite optical transitions of hexcitons in the photoluminescence spectrum. 
\end{abstract}

\maketitle

Hydrogen-like bound states of photoexcited electron-hole pairs in semiconductors -- that is, excitons -- have been a focus of considerable study for more than half a century \cite{Elliott_PR57, Laude_PR71, RashbaBook, HaugBook, Combescot_Book}.  In undoped direct-gap semiconductors, neutral excitons comprise the photogenerated electron and hole in the conduction and valence bands (CB and VB), respectively, and typically manifest as discrete optical resonances below the free-particle band-gap energy. More interesting states arise when electron-hole (\textit{e-h}) pairs are photoexcited into doped semiconductors containing a Fermi sea of mobile carriers \cite{Mahan_PR67b, Chang_PRB85, Sooryakumar_PRL87, Skolnick_PRL87, Haug_PQE, HaugBook, Combescot_Book, Hawrylak_PRB91, Kane_PRB94}.  Most well-known are the charged excitons (or trions) that emerge in lightly electron- or hole-doped semiconductors \cite{Stebe_PRL89, Kheng_PRL93, Finkelstein_PRL95}.  In the simplest picture, a trion is a three-particle complex consisting of a carrier from the Fermi sea bound to the photoexcited \textit{e-h} pair. Trions appear as an additional optical resonance below the neutral exciton, and the energy difference between the two provides a measure of the binding energy between the exciton and the resident carrier: $\sim\,$1-5~meV in semiconductors such as GaAs or ZnSe \cite{Huard_PRL00, Bracker_PRB05, Astakhov_PRB00}, and $\sim\,$20-30~meV in monolayer (ML) semiconductors such as MoS$_2$ or WSe$_2$ \cite{Wang_RMP18}.

A proper microscopic description of charged excitons remains a topic of active debate and long-standing theoretical interest, rejuvenated recently by detailed spectroscopy of charge-tunable transition-metal dichalcogenide MLs \cite{Wang_NanoLett17, Smolenski_PRL19, Wang_PRX20, Liu_PRL20, Liu_NatComm21, Li_NanoLett22, Wang_RMP18}.  For example, drawing on concepts in cold-atom physics \cite{Schmidt_RPP18, Massignan_RPP14}, it was recently suggested that the optical resonances attributed to trions and excitons can be regarded as the attractive and repulsive branches, respectively, of the collective Fermi-sea response to a photoexcited \textit{e-h} pair \cite{Sidler_NatPhys17, Efimkin_PRB17, Efimkin_PRB18, Fey_PRB20}.  An alternative picture, dating back over two decades, considers the ground state as a \textit{four}-particle ``tetron'' complex, wherein the trion moves together and is correlated with the hole in the Fermi sea (the so-called Fermi or CB hole) that is created when the trion forms \cite{Bronold_PRB00,Suris_PSS01,Esser_pssb01,Koudinov_PRL14}. Relationships between these various pictures have been discussed in recent literature \cite{Rana_PRB20, Efimkin_PRB21, Glazov_JCP20,Chang_PRB18}. Common to these pictures is that the photoexcited \textit{e-h} pair interacts primarily with resident carriers that are quantum-mechanically distinguishable. The electrons can then stay together near the VB hole without violating the Pauli exclusion principle. In electron-doped semiconductors such as GaAs or MoSe$_2$, photoexcited electrons belong to one of the two reservoirs populated by resident electrons. As a result, the  \textit{e-h} pair primarily interacts with one available reservoir of electrons with distinguishable quantum numbers.

\begin{figure*}
\includegraphics[width=17cm]{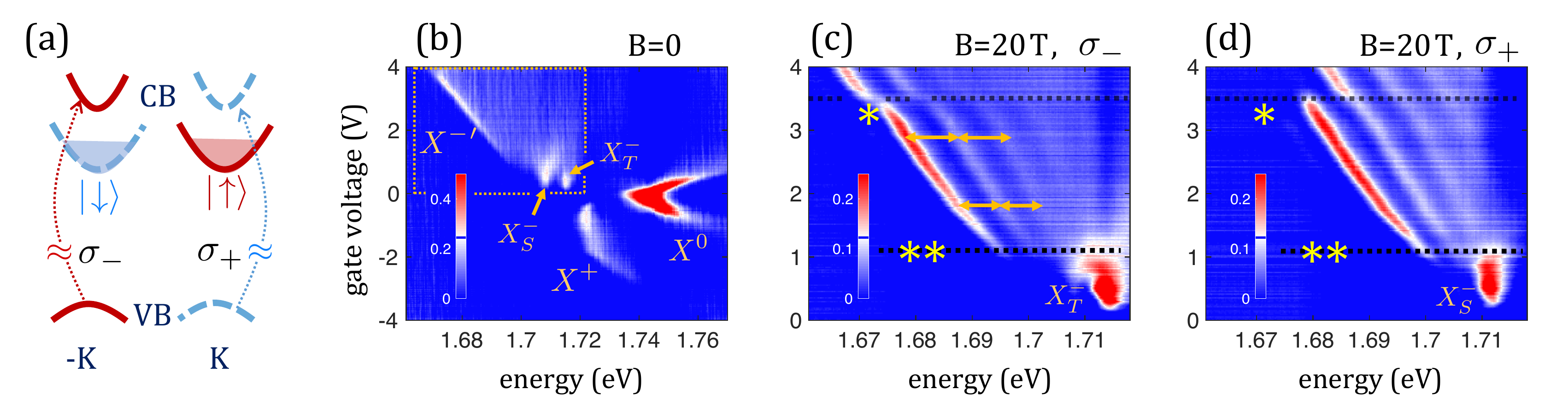}
 \caption{(a) Helicity resolved optical transitions in  ML-WSe$_2$. Light excitation with $\sigma_+$ ($\sigma_-$) helicity corresponds to  optical transitions  at the $K$ ($-K$) valley. Resident electrons occupy the bottommost valleys, whereas photoexcited electrons belong in the top valleys. (b) Optical reflectance spectra at 4~K as a function of gate voltage and photon energy. (c) and (d) Helicity resolved magneto-optical reflectance spectra  when the out-of-plane magnetic field is 20~T, shown in the spectral and voltage windows that are marked by the dotted box in (b).} \label{fig:1}
\end{figure*}

However, entirely new classes of increasingly complex many-body excitons are predicted to emerge if the photoexcited \textit{e-h} pair can interact simultaneously with \textit{multiple} distinguishable electrons \cite{s,l}. Fortunately, the ordering of the spin-polarized CB valleys at the $\pm K$ points of ML-WSe$_2$ provides an ideal platform to investigate such novel many-body states. As depicted in Fig.~\ref{fig:1}(a), the low-energy excitonic transition in this ML couples to the \textit{top} (not bottom) CB valleys at the $\pm K$ points. This leaves the photoexcited \textit{e-h} pair free to interact with both reservoirs of distinguishable electrons in the bottommost CB valleys. In this Letter,  we present magneto-optical experiments that provide evidence for 6-particle ``hexcitons'' and even 8-particle ``oxcitons'' that emerge when a photoexcited \textit{e-h} pair in doped WSe$_2$ interacts simultaneously with two or even three reservoirs of distinguishable electrons. Numerical calculations of the hexciton binding energy support this scenario \cite{g}.

Figure~\ref{fig:1}(b) shows the density-dependent optical reflectivity spectra from a charge-tunable single ML-WSe$_2$ at zero magnetic field (see Ref.~\cite{Wang_PRX20} for sample and experimental details).  Essentially, similar behavior has been reported by several groups \cite{Wang_NanoLett17,Liu_NatComm21, Li_NanoLett22}. When the ML is undoped, only the neutral exciton resonance appears ($X^0$).  Positively- and negatively-charged excitons ($X^\pm$) emerge at lower energies when the ML is lightly hole- and electron-doped, respectively. Note that \textit{two} distinct $X^-$ resonances appear in WSe$_2$, because photoexcited \textit{e-h} pairs can interact with either opposite-spin electrons in the same valley (singlet, $X^-_S$), or with same-spin electrons in the opposite valley (triplet, $X^-_T$), forming intravalley and intervalley trions \cite{Jones_NatPhys16,Courtade_PRB17}.   

At higher electron density $n_e$, the features $X^-_{S,T}$  disappear and a strong new resonance, often called $X^{-\prime}$, emerges at even lower energy and redshifts further with increasing $n_e$. Its origin is not understood, despite having been first observed in 2013 \cite{Jones_NatNano13}. That a similar resonance never appears on the hole-doped side is already a compelling argument that $X^{-\prime}$ is not related to simple exciton-polarons, tetrons, or transitions between free electron-hole pairs. We argue below that $X^{-\prime}$ corresponds to the stable formation of a 6-body hexciton state, which forms in ML-WSe$_2$ when a photoexcited \textit{e-h} pair interacts simultaneously with electrons from the bottommost CB valleys at $K$ and $-K$. 

We focus on the emergence and evolution of the optical transition with charge density in applied magnetic fields $B$. The field breaks the valley degeneracy, thereby allowing us to better control the population in each of the four spin-polarized CB valleys. Figures~\ref{fig:1}(c) and (d) show density-dependent maps of the $\sigma^\pm$ circular-polarized optical reflectivity at $B$=20~T.  At this large field, the resonance splits to a series of lines that are reminiscent of Landau level (LL) formation. 

\begin{figure}
\includegraphics[width=8.2cm]{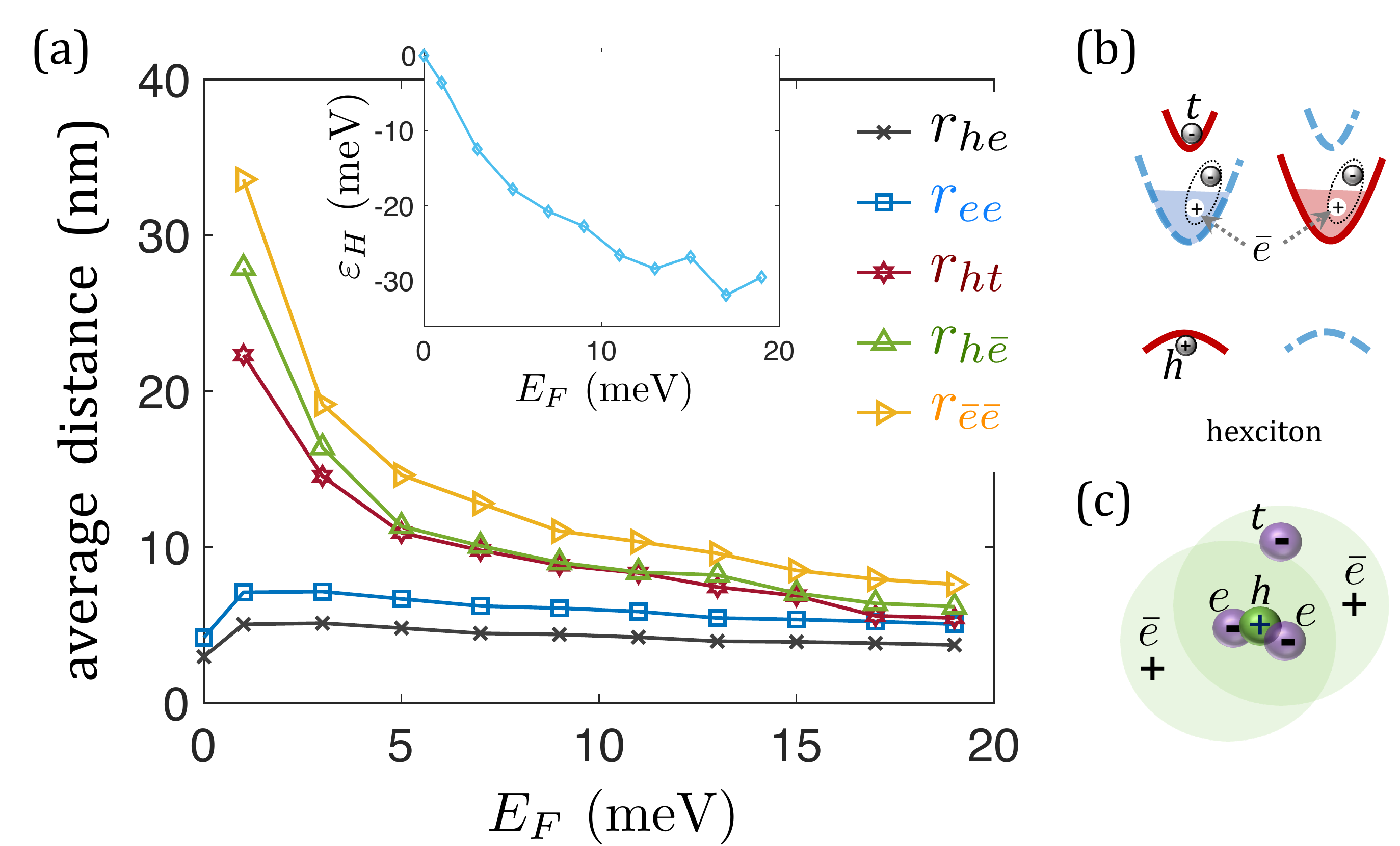}
 \caption{(a) Calculated inter-particle distances in the hexciton as a function of the Fermi energy ($E_F \approx 5$~meV amounts to $n = 10^{12}$~cm$^{-2}$). (a) Results are shown for average distances between the CB electron and VB hole ($r_{he}$), between the two CB electrons of the core trion ($r_{ee}$), between the VB hole and the outer top-valley electron ($r_{ht}$) or CB holes ($r_{h\bar{e}}$), and between the CB holes ($r_{\bar{e}\bar{e}}$). Inset: Calculated binding energy of the satellite electron to the hexciton. (b) and (c) Schemes of the hexciton in $k$-space and real-space, respectively.  The trion at the core of the hexciton binds to two CB holes and a satellite electron.} \label{fig:calc}
\end{figure}

A first piece of evidence favoring hexciton formation is that the LLs appear \textit{simultaneously} in both $\sigma^\pm$ polarizations when $V_g \approx 1$V (corresponding to $n_e \sim 1-2\times10^{12}$~cm$^{-2}$). At smaller densities, free electrons fill only the spin-up CB valley at $K$, whose energy is shifted down by the magnetic field.  Consequently, the spectrum only includes the resonances $X_S^-$ with $\sigma_+$ polarization and $X_T^-$ with $\sigma_-$ polarization. At voltages larger than $\sim1$~V, indicated by dotted lines with two asterisks in Figs.~\ref{fig:1}(c) and (d), the Fermi sea begins to fill the spin-down CB valley at $-K$.  This threshold provides two distinguishable reservoirs of free electrons with which a photoexcited \textit{e-h} pair can simultaneously interact, forming the 6-particle hexciton (depicted in Figs.~\ref{fig:calc}(b) and (c)).

\begin{figure}
\includegraphics[width=8.5cm]{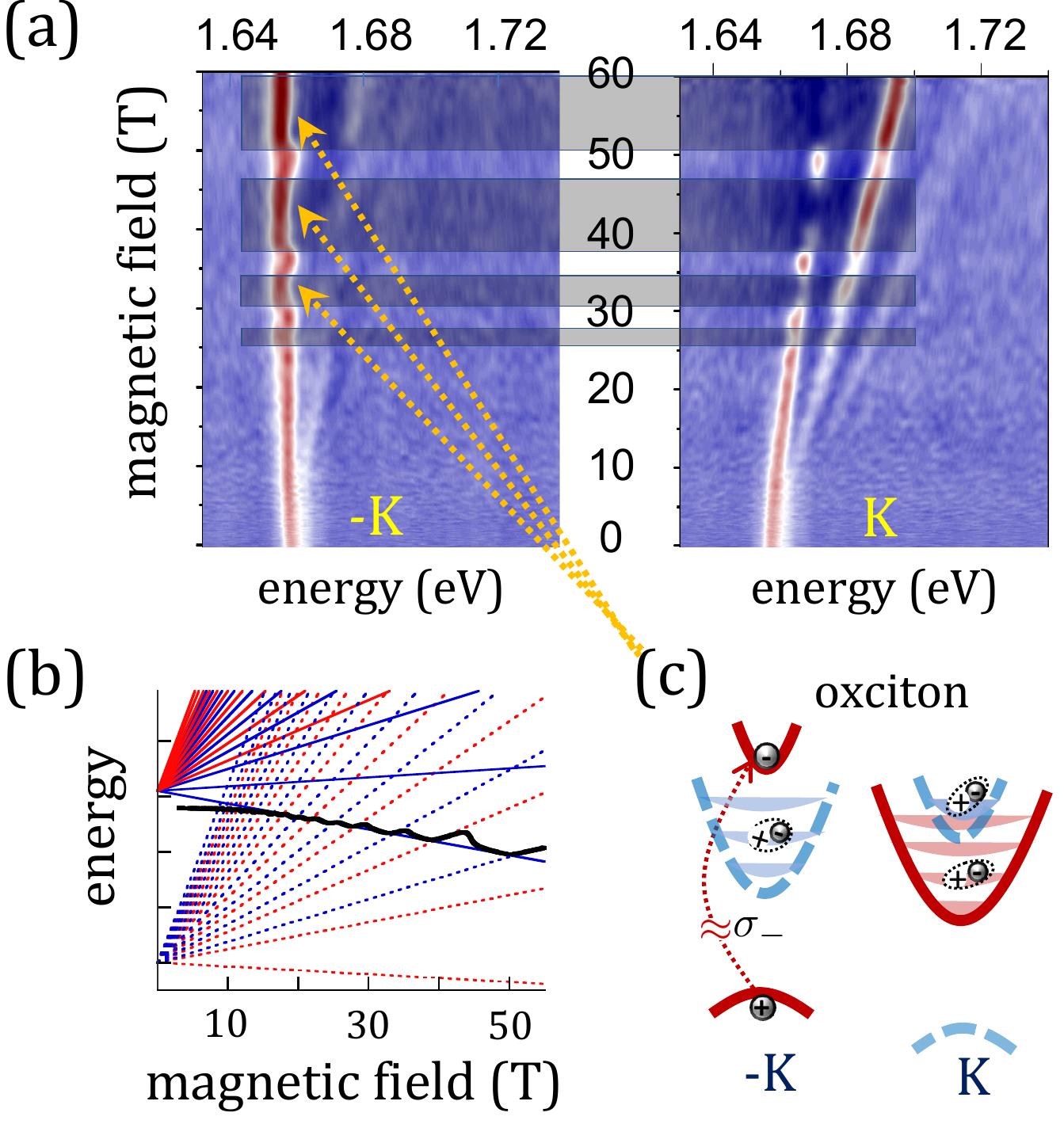}
 \caption{ (a)  Magneto-optical reflectance spectra as a function of magnetic field and  photon energy in a second device. The voltage level is at the threshold of filling the top CB valley of $K$. The shaded boxes show the regimes at which the top valley at $K$ is filled. (b) Spin- and valley-resolved LL diagrams in the CBs within a single-electron picture. The solid black line denotes the chemical potential of the Fermi sea. (c) The oxciton state, formed when the trion at its core binds to three Fermi holes in the CB and two satellite electrons. The hexciton-to-oxciton transition takes place when the top CB valley at $K$ has one filled Landau level, marked by the shaded regions in (a).} \label{fig:2}
\end{figure}

A second noteworthy feature occurs at larger electron densities, when $V_g \approx$3.5~V, indicated by dotted lines with one asterisk in Figs.~\ref{fig:1}(c) and (d). The lowest energy resonance with $\sigma^+$ polarization abruptly disappears, indicating that the Fermi sea has just filled the lowest LL in the top CB valley at $K$, which quenches the corresponding optical transition due to Pauli blocking. More importantly, concurrent with the quenching in $\sigma_+$, the optical resonances in $\sigma_-$ exhibit a discontinuous redshift. Namely, the binding energy of the complex in $-K$ abruptly \textit{increases} (by $\sim$ 3meV) when free electrons fill not only the two bottommost CB valleys but also a third CB valley having distinguishable quantum numbers (spin-down in $K$).  This threshold marks the transition from hexcitons to oxcitons, schematically shown in Fig.~\ref{fig:2}(c).

These spectroscopic signatures are also observed in complementary magneto-absorption studies of another charge-tunable ML-WSe$_2$, shown in Fig.~\ref{fig:2}. Here, $n_e$ is fixed and $B$ is swept from 0-55~T in a pulsed magnet (see Ref.~\cite{Li_NanoLett22} for sample and experimental details).  When $n_e$ is set just below the point where electrons begin to fill the top CB valleys (at zero field), then the combination of valley Zeeman splitting and LL formation forces the upper CB in $K$ to repeatedly fill and empty as $B$ increases. Figure~\ref{fig:2}(b) shows the calculated chemical potential of the Fermi sea (black solid line), superimposed on the LL fan diagram. Repeated Pauli blocking and unblocking of the  resonance in $K$ is observed. Concomitantly, the resonance in $-K$ exhibits an abrupt redshift of $\sim$3~meV whenever electrons occupy the third CB valley, indicating an abrupt increase in the binding energy of the excitonic state. As with the gate-dependent studies in Figs.~\ref{fig:1}(c) and (d), this behavior is consistent with the hexciton-to-oxciton  transition, occurring when the number of distinguishable electron reservoirs (with which the photoexcited e-h pair in $-K$ can interact) increases from two to three.  

We emphasize that the observed discontinuities of the resonance as a function of $n_e$ and $B$ cannot be explained within a picture of transitions between free-particle LLs in the CB and VB, and are not consistent with our own previous claim for coupling between excitons and intervalley plasmons \cite{VanTuan_PRX17,Dery_PRB16}. The Supplementary Material (SM) provides further evidence against the possibilities that neither intervalley plasmons nor  optical transitions between LLs of free electron-hole pairs can stand behind the observed phenomena. 

A third spectroscopic feature supporting hexcitons is the LL separation $\hbar \omega_c = \hbar e B / m^*$, where $\omega_c$ is the cyclotron frequency and $m^*$ is the effective mass. As indicated by the arrows in Fig.~\ref{fig:1}(c), the LL separation increases with gate voltage, revealing that $m^*$ falls from $\sim0.32m_0$ to $\sim0.26m_0$ as $n_e$ increases ($m_0$ is the free-electron mass). These masses are evidently larger than the reduced mass of  neutral excitons in WSe$_2$ \cite{Stier_PRL18, VanTuan_PRB18}, and are much closer to the electron mass in the top CB valleys \cite{Kormanyos_2DMater15}. Thus, the LLs are likely related to the quantized motion of the top-valley electron. We explain how this behavior, and also the redshift of the LLs with increasing $n_e$, are consistent with optical transitions of hexcitons.

Figure~\ref{fig:calc} shows calculated average inter-particle distances within the hexciton as a function of the electron density.  Details of the theory needed for these calculations are presented in Refs.~\cite{s,l}, and here we discuss the results in the experimental context. $r_{he}$ is the calculated average distance between the VB hole and electron in the bottommost CB valley. $r_{ee}$ denotes the respective distance between the two electrons in the bottommost CB valley. Compared with the other shown cases, the small values of  $r_{he}$  and $r_{ee}$ imply that the trion at the hexciton core is dark, comprising the optically inactive bottom-valley electrons. The VB hole prefers binding tightly to these two electrons on accounts of their heavier mass compared with that of the photoexcited electron in the top valley \cite{Yang_PRB22,VanTuan_PRB18,Kormanyos_2DMater15}. As depicted in Fig.~\ref{fig:calc}(c), the latter behaves as a `satellite', capturing the electron-depleted region around the core trion, rendered by the two CB holes. As evident from the distance between the satellite electron and the VB hole ($r_{ht}$),  the binding of the former to the complex is relatively weak at small electron densities, as shown by the inset of Fig.~\ref{fig:calc}. The spatial extent of the CB holes is large at small densities, as can be understood from the average distances between the VB and CB holes ($r_{h\bar{e}}$) or between the CB holes ($r_{\bar{e}\bar{e}}$). Connecting these results with the experimental findings, the effective mass we extract from the energy difference between LLs  in Figs.~\ref{fig:1}(c) and (d) is nearly that of a top-valley free electron at small electron densities (i.e., when its binding to the complex is small). The energy difference between LLs increases when the density increases, indicating of a smaller effective mass. This behavior implies that the motion of the satellite  electron and core dark trion becomes correlated, caused by tighter binding of the former to the hexciton, as shown in Fig.~\ref{fig:calc}. The effective mass in this case would ultimately correspond to the reduced mass, $1/m^* = 1/m_t + 1/M_D$ where $M_D$ is the translational mass of the dark trion ($M_D \sim 4m_t$ in ML WSe$_2$ \cite{Yang_PRB22}).

\begin{figure}
\includegraphics[width=8.5cm]{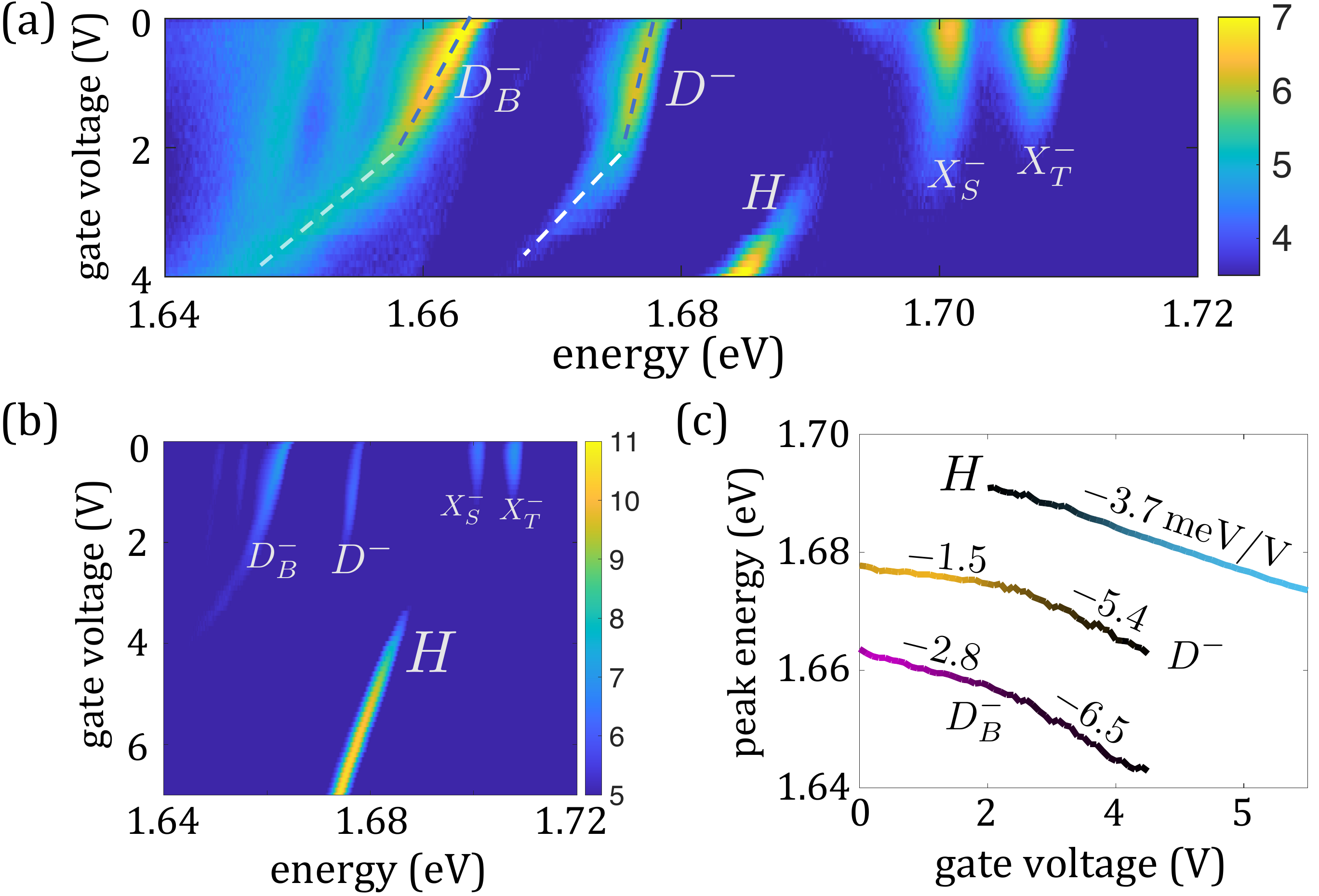}
 \caption{ (a) Log-scale colormap of the photoluminescence intensity in electron-doped ML-WSe$_2$ at 4~K.  The hexciton peak, marked with $H$ (often dubbed $X^-\,\!'$ in the literature), emerges around 2~V. The secondary transitions emerge from the dark trion optical transitions, $D^-$ and $D^-_B$ (see text), and their redshift is enhanced when the peak $H$ emerges. (b) The colormap in a larger voltage range, showing the dominance of $H$ in the electron-rich regime. (c) Peak energies of $H$, $D^-$ and $D^-_B$. The line brightness denotes the normalized PL intensity of the peak and the redshift numbers are in meV per Volt.} \label{fig:PL}
\end{figure}

Additional support for the presence of hexcitons can be found in the photoluminescence (PL) spectrum. Figure~\ref{fig:PL} shows the  low-temperature PL intensity at zero magnetic field from a third-charge tunable ML-WSe$_2$  device. Optical transitions of this device close to charge neutrality were analyzed in Ref.~\cite{He_NatComm20}. Here, we focus on the dominant transition when the ML is electron rich, marked with $H$ (hereafter we refer to $X^-\,\!'$ as $H$). Its PL intensity is noticeable when the gate voltage is larger than $\sim$2~V. If the hexciton picture is correct, then the energy of the peak $H$ should  emerge from $\hbar\omega_{D^-} + \Delta_c$ at small electron densities, where $\hbar\omega_{D^-}$ is the optical-transition energy of the dark trion and $\Delta_c \simeq 14$~meV is the spin splitting energy between the bottom and top CB valleys \cite{Kapuscinski_NatComm21}.  To understand this reasoning, we recall that the binding energy of the hexciton is that of the dark trion at its core when the spatial extent of the CB holes is large (small density), which in turn means that the outer (top-valley) electron is loosely bound to the complex. Yet, the energy emerges from  $\hbar\omega_{D^-} + \Delta_c$ rather than  $\hbar\omega_{D^-}$  because the optical transition involves the tightly bound VB hole and loosely bound top-valley electron. When the electron density increases, the spatial extent of the CB holes shrinks and the binding of the outer electron to the complex increases, as shown in Fig.~\ref{fig:calc}. Hence the observed redshift and amplified PL. 

We explain the observed behavior in Fig.~\ref{fig:PL}  as follows. Hexcitons correspond to three optical transitions marked with $H$, $D^-$ and $D^-_B$  when the electron density is  large. Each of these transitions correspond to the recombination of a different electron in the complex with the VB hole. At small electron densities, shown by the voltage range $\lesssim$2V  in Fig.~\ref{fig:PL}(a), the outer top-valley electron is still far from the complex, resulting in vanishing PL intensity of the peak $H$. Despite having `wrong' quantum numbers, the PL intensities of the dark trion and its brightened component are stronger than $H$ when the electron density is small: $D^-$ ($D^-_B$)  involves recombination of the electron with opposite spin (valley) compared to that of the missing electron in the VB. The peak $H$ emerges gradually when the electron density continues to increase;  a result of the tighter binding of the top-valley electron to the hexciton. With the emergence of $H$, the redshifts of $D^-$ and $D^-_B$ are enhanced, accompanied with their gradual decay.  Figure~\ref{fig:PL}(c) shows the peak energies of $H$, $D^-$ and $D^-_B$, where the brightness of the line corresponds to the normalized PL intensity of the peak. The redshifts of $D^-$ and $D^-_B$ are enhanced around  $\sim$2~V,  where the slope of $D^-$ changes  from $-1.5$ to $-5.4$~meV/V, and that of $D^-_B$ from $-2.8$ to $-6.5$~meV/V. The differences match nearly perfectly with the slope of $H$ ($-3.7$~meV/V). The density regime at which the slopes of $D^-$ and $D^-_B$ are enhanced mark the transition from trions to hexcitons. Consistent with that,  the emergence of $H$ in  Fig.~\ref{fig:PL}(a) is also accompanied with fading of the bright trions  ($X^-_{S,T}$). The SM includes further analysis on the redshifts of $H$, $D^-$ and $D^-_B$. 

In conclusion, we have provided supporting evidence for the presence of hexcitons and oxcitons in electron rich ML-WSe$_2$. While this achievement is important on its own merits, lest we lose sight of the wider implications. Among the various mechanisms proposed in the literature so far to explain optical transitions of bound excitonic states in doped semiconductors, our findings reinforce the viability of composite excitonic states \cite{l,s}. The unique band structure of ML-WSe$_2$ allows us not only to reinforce this idea but to provide the following general principle. When the exciton binding energy of the semiconductor exceeds the Fermi energy, bound excitonic states are made of one or more electron-hole pairs, where the hole of at least one pair is from the VB. Other electron-hole pairs are made of CB holes and electrons with distinct spin-valley configuration. The possible number of pairs in a composite excitonic state is determined by the spin-valley space at the edge of the CB (or VB by reversing the discussion to $p$-type doping). 

In experiment, composite excitonic states are seen when the charge density is such that the spatial extent of CB holes is not far larger than the bare-trion radius. In this regime, CB holes can further glue extra electrons around the core trion. Many semiconductors, including zinc blende compounds, can host up to two electrons in a tetron-type configuration \cite{g}. Other semiconductors can host larger composites, such as the hexciton or oxciton in ML-WSe$_2$, and possibly beyond that in multi-valley semiconductors. The observation of such many-body correlated complexes can be facilitated by engineering the dielectric environment to enhance the binding energies. Using magnetic fields or strain to lift the spin and valley (pseudospin) degeneracies, one can further control the size of the excitonic state.

\acknowledgments{Dinh Van Tuan is supported by the Department of Energy, Basic Energy Sciences, Division of Materials Sciences and Engineering under Award DE-SC0014349. Hanan Dery is supported by the Office of Naval Research, under Award N000142112448. The work at NHMFL (Scott A. Crooker) is supported by NSF DMR-1644779, the State of Florida, and the  Department of Energy. Xiaodong Xu is supported by Department of Energy, Basic Energy Sciences, under award DE-SC0018171. Su-Fei Shi is supported by NSF, under Career award DMR-1945420 and award DMR-2104902.}

\end{document}